\documentstyle[12pt]{article}
\def\beq{\begin{eqnarray}}   \def\eeq{\end{eqnarray}}
\def\d{\delta(y)}
\def\a{{\cal A}}

\begin{document}
\begin{flushright}
NYU-TH/00/09/09 \\
TPI-MINN-00/50\\
UMN-TH-1926/00
\end{flushright}

\vspace{0.1in}
\begin{center}
\bigskip\bigskip
{\large \bf (Quasi)Localized Gauge Field on a Brane: 
\\
\vspace{0.2in}
Dissipating Cosmic Radiation to Extra Dimensions?}

\vspace{0.5in}      

{Gia Dvali$^1$, Gregory Gabadadze$^2$, and M. Shifman$^2$ }
\vspace{0.1in}

{\baselineskip=14pt \it 
$^1$Department of Physics, New York University,  
New York, NY 10003 \\
$^2$ Theoretical Physics Institute, University of Minnesota, 
Minneapolis, MN 55455}  
\vspace{0.2in}
\end{center}

\vspace{0.9cm}
\begin{center}
{\bf Abstract}
\end{center} 
\vspace{0.1in}

We propose  a mechanism ensuring (quasi)localization of massless
gauge fields on a brane. The mechanism does not rely
on BPS properties of the brane 
and can be realized in any theory where charged particles
are confined to the world-volume.
The localized  matter fluctuations induce 
a gauge kinetic term on the brane. At short distances
the resulting propagator for the gauge
field is {\it identical} to the   
four-dimensional propagator. The gauge theory on the brane is
effectively four-dimensional at  short distances;
 it  becomes higher-dimensional on very large (cosmic) scales. The
brane-bulk system exhibits the  phenomenon of ``infrared
transparency''.  As a result, only very low frequency modes  can
escape into extra dimensions. In this framework the large wavelength
cosmic radiation can dissipate in  extra space  at a rate that may
be observable, in principle. We briefly discuss some astrophysical
consequences of this scenario. 

The same mechanism of localization of gauge fields may
work in  Kaplan's framework for domain
wall chiral fermions on lattices.

\newpage

\section{Introduction}

The idea that the Standard Model particles can be localized on a brane
may dramatically change the low-energy implications of extra
dimensions and quantum gravity. In conjunction with simple
compactifications, this idea allows for extra dimensions to be
macroscopically large ($\sim 10^0 $  mm or so), with the fundamental
quantum gravity scale being as low as a few TeV, thus providing an
alternative view on the hierarchy problem \cite {ADD}. Noncompact,
but finite-volume extra dimensions are also possible if one introduces a
negative bulk cosmological constant \cite {RandallSundrum}.

Putting aside the hierarchy problem, the
brane world idea may have other implications, e.g.,
it can give rise to a natural mechanism for supersymmetry breaking.
If the brane at hand is a non-BPS state then
it breaks  all  supercharges  of the bulk space
rendering the four-dimensional world free of the  Fermi-Bose
degeneracy \cite {GiaMisha}. Moreover, one may naturally explain in this
way the smallness of the SUSY-breaking parameters \cite{FAN}.                
     
In the brane world model-building the question
 of paramount importance is the localization
of the Standard Model particles on the brane. A mechanism of the fermion
localization was suggested long time ago in \cite{RubakovShaposhnikov},
based on an index theorem of Ref. \cite {JackiwRebbi}.
The localization mechanisms for scalar and fermion 
fields in  warped backgrounds are known as well \cite{BorutGiga}
(see also \cite {RandjbarShaposhnikov}).

The vital  issue is how vector fields can be
localized on the brane. This question will be the
main concern of the present paper.
The gauge fields can  be localized on D-branes very naturally
in the  framework of  string theory \cite {Polchinski}. D-branes are
BPS objects that preserve half of the original supersymmetry;
their low-energy world-volume theory
is described by massless fluctuations of open strings ending
on them. In fact, the localized vector fields on a D-brane
world-volume are just spin-1 massless fluctuations of the
corresponding open strings. Unbroken supersymmetry then guarantees that
these spin-1 fluctuations stay massless in order to correctly fill up the
supermultiplets together with the massless world-volume scalars 
and fermions. 

However, in the brane world context it is highly desirable 
that the brane is not a BPS state  since supersymmetry 
must be broken in a world-volume theory. 
Whether or not the same mechanism of localization of
the vector fields
holds for non-BPS branes, especially those which cannot be presented as
bound states  of BPS
states, is not clear. In this respect it is crucial  to develop 
additional field-theoretic mechanisms for the
massless spin-1 localization on a  brane,
which would be independent of BPS properties.

One such mechanism was proposed in Ref. \cite {GMLoc}, where 
the gauge theory is assumed to be confining in the bulk but not 
on the brane. As a result the electric charges on the brane are not 
screened, but rather are in a Coulomb phase since 
the electric flux cannot spread in the bulk. 
Thus, there is a massless gauge field trapped on the brane
world-volume theory. In this case neither photon nor electric
charges can escape to extra dimensions.
Another possibilities  were  recently proposed  in
Refs. \cite {Oda} and \cite {Tinyakov} 
where the gauge fields were localized on the brane
in extra-dimensional space which was warped (and in 
addition compactified in \cite {Tinyakov}). 

In this work we will present a mechanism of
(quasi)localization of gauge fields on a brane which is
rather model-independent. This mechanism can be
used  in theories with compact, warped and, most importantly,
infinite-volume extra dimensions. This is a generalization of the 
mechanism of Ref. \cite {DGPind} for the localization of a
massless spin-2 field on the brane  due to a ``brane-induced''
world-volume Ricci term. The main idea of  Ref. \cite {DGPind} is as 
follows:  Consider a bulk field coupled minimally to some of the
brane-localized degrees of freedom. Then, a four-dimensional
kinetic term for a given bulk field is  generated  on the brane
world-volume due to loop corrections associated with
localized matter fields \cite {DGPind}. This brane-induced  kinetic term
ensures that the propagation of the bulk field on the brane 
is {\it four-dimensional} at the
distances smaller than a certain crossover scale $r_c$.
In this way brane ``induces'' an effectively four-dimensional theory
in its world-volume.
In Ref. \cite{DGPind} this idea was used for the (quasi)localization of the
massless spin-2 field on a brane.
In this case the four-dimensional laws of gravity can
be obtained on a {\it delta-function} type brane
even if extra dimensions have  infinite  volume \cite {DGPind,DG}.   
This mechanism is quite universal since it requires only the
presence of localized spinor and/or scalar fields on the brane.
Moreover, the same mechanism is valid even if there are 
no localized matter fields on the brane but the brane has a finite
thickness. In the latter case the four-dimensional Ricci scalar is
induced in the low-energy world-volume theory after  the effects of the 
finite brane thickness are integrated out \cite {DG}. 
In this framework  there is no need to compactify or to warp the 
extra space, it can have an infinite volume \cite {DGPind,DG}.

In the present work we will generalize the above mechanism
for spin-1 fields.
For this mechanism to work it is sufficient to have bulk gauge fields
and a brane on which charged matter fields are localized. Then the
quantum fluctuations of localized matter automatically take
care of (quasi)localization
of  the bulk gauge fields.
To be specific, in the present work we deal with five-dimensional models.
Higher dimensions can be treated in a similar manner.
In Sect. 2, we consider an idealized case
of a five-dimensional theory with a
delta-function  type brane. This simplification is temporary
as in Sect. 3 we consider a field-theoretic setting
with a finite-thickness brane
\footnote{In those cases when there is a menace of
confusion,
we will refer to the field-theoretic setting
with the  finite-thickness brane as to a  domain wall, or wall for short,
reserving the term ``brane" for the object which has zero transverse width in
a classical approximation.}.   
The field-theoretic consideration
is  instructive  since  it allows us to deduce
an exact expression for a tree-level  four-dimensional potential
mediated by the quasilocalized gauge fields. Moreover, we note that
quasilocalization on the zero-thickness brane can be applied  to   objects
which   can be treated as singular sources in the weak coupling approximation
of string theory. Such are D-branes, orientifold planes
and/or  extended objects which are stuck to the orbifold fixed
points\footnote{This statement needs some qualifications
since  in Type IIA,B string theories in ten dimensions there are no
bulk gauge fields in the first place.
However, bulk gauge fields can emerge
once some of the extra six dimensions are compactified. For instance,
in compactification to five dimensions
on $K3\times S^1$  there are points of enhanced gauge
symmetry on the moduli space where the gauge fields emerge. Thus, we
can have a brane with localized matter as well as the bulk
 populated by the gauge fields.}
(all at weak coupling).         

In Sect. 3  we consider a five-dimensional field theory
with a domain wall soliton in which
we {\it derive} the results of Sect. 2 in the case of a finite-thickness
wall. In fact, we will show that
the domain wall  with the localized fermion zero-modes
produces a four-dimensional kinetic term for the gauge fields and,
thus, leads to quasilocalization of spin-1 particles.
 
Finally, in Sect. 4 we analyze
some phenomenological consequences  of quasilocalization.
Since now the gauge fields can escape in  the bulk,
local charges in the world-volume are not
necessarily conserved \cite {Tinyakov}.
A very  intriguing feature of the framework is that 
the large wavelength cosmic radiation
can dissipate into  the bulk over the cosmic distances, by the rate that
is, in principle, detectable. This offers a number of exciting astrophysical
implications for the above scenario, which we shall briefly discuss.

\section{Induced Gauge Fields on a Brane}

In this section we consider a simple case --  a brane of the
{\it delta-function}
type   embedded in five-dimensional space.
In other words, we will disregard effects due to
non-vanishing thickness of the wall.

We will not be 
interested in effects due to gravity. 
We rather assume  that the brane at hand is made of some heavy fields
which are already  decoupled from the low-energy dynamics 
(an explicit field-theoretic realization will be give in the next section). 
 
Suppose that there are some  gauge 
fields ${\cal A}_C$ (here $C=0,1,2,3,5$)   living
in the bulk  
for which the Lagrangian takes the form 
\beq
-{1\over 4g^2}~{\cal F}_{AB}^2~+{\rm other~~fields},
\label{bulk}
\eeq
where $g$ is a coupling constant with the dimensionality
\beq
[g^2]~=~[{\rm mass}]^{-1}~.
\label{dimg}
\eeq
For simplicity we can assume that ${\cal A}_C$ is an Abelian field;
similar considerations can be carried out in the
non-Abelian case.
The three-brane on which matter is localized can be put  
at the  point $y=0$. 

The Dirac-Nambu-Goto action for a brane takes the form:
\beq
S_{\mbox{3-brane}}\, =\, -T 
~\int d^{4}x~\sqrt{|{\rm det} {\bar g}|}\, ,
\label{brane0}
\eeq 
where $T$ stands for the brane tension,  and
$${\bar g_{\mu\nu}}=\partial_\mu X^A\partial_\nu X^B G_{AB}$$ 
denotes  the induced metric on the brane (with $G_{AB}$ being the metric of
five-dimensio\-nal space-time). 
$X^A$ ($A=1,2,...,5$) are the coordinates in 5D space. If for simplicity   
we neglect brane's  fluctuations (these are just scalar particles
which can easily be included in the consideration if so desired) then
$\partial_\mu X_5=0$, and $X_\mu=x_\mu$ so that the  induced metric 
can be written  as follows:
$${\bar g}_{\mu\nu}~(x)\, =\, \left. G_{\mu\nu}\left (x, y \right
)\right|_{y=0}\,.$$ 
In general, there could  be localized matter fields on the
brane  world-volume. One can    take them  into account by writing the 
following action:
\beq
{\tilde S}^{\rm matter}_{\mbox{3-brane}} =\,
S_{\mbox{3-brane}} +\, 
\int d^{4}x~\sqrt{|{\rm det} {\bar g}|}~{\tilde {\cal L}} (\psi)\, ,
\label{brane1}
\eeq  
where $\psi$ denotes collectively all   localized 
fields for which the four-dimensional 
Lagrangian density is ${\tilde {\cal L}}$. 
The world-volume field theory may be regarded as
an effective field theory with some cutoff $\Lambda$.
The current of the localized matter fields 
on a brane can  be written as follows:
\beq
J_A (x,y)\, =\, J_\mu (x)~\d~\delta^{\mu}_ A\, .
\label{current}
\eeq 
The five-dimensional current conservation associated
with the five-dimensional gauge invariance, implies
the four-dimensional current conservation on the brane, since  the fifth
component of the current vanishes.
 This current couples to the bulk gauge field. The interaction 
Lagrangian is
\beq
{\cal L}_{\rm int}\, =\, \int d^4x\, dy\, J_C(x,y)\, \a^C(x,y)
= \int
d^4x\, J_\mu(x)~ \a^{\mu}(x, 0)\, .
\label{int}
\eeq
Thus, the current of localized matter interacts with 
the  gauge field 
\beq
A_\mu(x)  \equiv  {\cal A}(x, y=0)\, .
\label{pull}
\eeq 
Due to the interaction (\ref {int})   an induced kinetic
term for the field $A_\mu(x)$ emerges on the brane world-volume. 
This term is generated by a one-loop diagram 
with two external legs of $A_\mu(x)$ and   localized matter
 running in the loop. The four-dimensional current conservation
implies transversality of this loop;
as a result,  
the following term should  be 
included in the total low-energy action on the brane
\beq
-{1\over {4 e^2}}\, F_{\mu\nu}^2 +\, {\rm higher~derivative~terms}\, ,
\label{4DF}
\eeq
where
$$
e^{-2} = \frac{2N_f}{3\pi}\ln \frac{\Lambda}{\mu}\,,
$$
$\Lambda$ and $\mu$ are the ultraviolet and infrared cut-offs, respectively.
  The sign of the induced term is  
negative, as it should be,
 since it is generated by localized scalar  and/or  spinor
fields  (confined to  the 
brane world-volume)    running in the  loops. Were the loops generated
by localized
vector fields, the sign would be wrong (positive).

The total (five-dimensional) Lagrangian for the gauge fields takes the form 
\beq
-{1\over 4g^2}\, {\cal F}_{AB}^2- {1\over {4 e^2}}\, 
F_{\mu\nu}^2\, \d + {\rm other~interactions}\,.
\label{total}
\eeq
Below we  will study the impact of the  induced kinetic term in (\ref {total})
at the tree-level.

Our  first task is to calculate the Coulomb potential induced by a 
probe charged
particle placed on the brane. To this end, we add a  source term 
${\cal A}^BJ_B$
to the Lagrangian (\ref{total}). 
The equation of motion in the theory  with  the source  is
\beq
{1\over g^2}~\partial_C\partial^C \a_B
+ \d\frac{1}{e^2}\left (  \partial_\mu\partial^\mu \a_\nu + \partial_\nu
(\partial_y \a_y)\right )\, \delta_{\nu B}\, =\, J_B(x,y)\, ,
\label{eqm}
\eeq
where we have chosen the Lorentz gauge in the bulk 
\beq
\partial^C \a_C = 0\, .
\label{lorentz}
\eeq

Next, we assume the source to be localized in the $y$ direction (see
Eq.  (\ref {current})) to transform  Eq. (\ref {eqm}) into
\beq
&&\partial_C\partial^C \, \a_{\mu}\, +\, \frac{g^2}{e^2} \d \, \left
[\partial_\beta\partial^\beta
\a_\mu~+~ \partial_\mu (\partial_y \a_y) \right ]= g^2\,  J_\mu(x) \d \, ,
\nonumber\\[0.2cm]
&&\partial_C\partial^C \, \a_y= 0\, .
\label{eqs}
\eeq
To derive the  propagator for the 
gauge field $A_\mu(x)$  it is convenient
to Fourier-transform these equations with respect to the  
four world-volume coordinates. The corresponding momenta  will be
denoted by $p$;  we will   work  in  Euclidean space, 
 $$p^2=p_4^2+p_1^2+p_2^2+p_3^2\,.$$
Equations  (\ref {eqs}) can be written as follows:
\beq
&&(p^2- \partial_y^2) {\tilde \a}_\mu (p,y) +\frac{g^2}{e^2}\, \d  
\left ( p^2 {\tilde \a}_\mu (p,y) +\, ip_\mu \partial_y {\tilde \a}_y(p,y)
 \right )\nonumber\\[0.2cm]&&= g^2\, {\tilde J}_\mu (p)\, \d\,,
\label{eq1} \\[0.3cm]
&&(p^2- \partial_y^2)\, {\tilde \a}_y(p,y)=0\,,
\label{eq2} 
\eeq
where the tilde denotes the Fourier transform.
Equation (\ref {eq1}) can be readily solved.  Multiplying both sides by
${\tilde J}_\mu (p)$ and using the four-dimensional transversality
(i.e. $p^\mu {\tilde J}_\mu (p)=0$) we obtain
\beq
{\tilde \a}_\mu(p,y) {\tilde J}^\mu(p) = e^2\left[{\tilde J}_\mu (p)\right]^2
\frac{1}{ p^2+ (2pe^2/g^2)}\, \exp \left (-p|y| \right ) 
,\qquad p\equiv\sqrt{p^2}
\,. 
\eeq
It describes   the gauge field   with somewhat unusual propagator, more of
which will be said later. 

Moreover, Eq. (\ref {eqm}) 
combined with the  four-dimensional transversality
of the current implies that $\partial^\mu {\cal A}_\mu (x,y) =0$,
which entails, in turn, from  the gauge fixing condition (\ref {lorentz})
that
\beq
\partial_y {\cal A}_y(x,y)=0\,.
\eeq
Combining this with 
 (\ref {eq2}) we get the following   equation  for the 
fifth component of the bulk gauge field:
\beq
\partial_\mu^2 {\cal A}_y(x,y)
=0\,.
\eeq
The latter equation describes   a four-dimensional massless scalar field  
which is decoupled from the matter fields localized on the brane
\footnote{Effects due to interactions of this scalar with 
the matter will be considered in 
the next section.}. 
Therefore, 
from the point of view of a four-dimensional   observer we can forget 
about ${\cal A}_y$ for the time being, and focus on
$A_\mu$. 

The propagator 
of the gauge field on the brane world-volume at $y=0$ 
takes the form 
\beq
D_{\mu\nu}(p) = {\eta_{\mu\nu} 
\over p^2+(2e^2 p/g^2)}\left[1+{\cal O}(p)\right]\,.
\label{prop}
\eeq
Physics of this type of propagators was elucidated in 
Ref. \cite {DGPind}.
There are two distinct regimes (a critical value
of momentum is 
$p_* =  {2e^2}/{g^2}$). 
 For small momenta, $p\ll p_*$  (i.e. at large distances) 
the propagator resembles that of the five-dimensional  theory. However, at 
large momenta, $p\gg p_*$ (i.e. at small distances) it behaves as  
the four-dimensional propagator. 
The crossover scale is determined by the 
bulk coupling constant $g^2$.

It is instructive to study these properties in more detail
in the coordinate space. For static probe charges only the time
component of the current survives; it is not difficult
to calculate the tree-level scalar potential 
due to the exchange of the  vector field  (\ref {prop}). 
Using the results of  Ref. \cite {DGPind} we get 
\beq
V(r)~\propto~{1 \over r}~\left \{ {\rm sin} 
\left ( {r\over r_c} \right ) ~{\rm Ci} \left ( {r\over r_c} \right )
~+~{1\over 2}  {\rm cos} 
\left ( {r\over r_c} \right ) \left 
[\pi~-~2 ~ {\rm Si} \left ( {r\over r_c} \right ) \right ]   \right \}~,
\label{V} 
\eeq
where $$ {\rm Ci}(z) 
\equiv \gamma +{\rm ln}(z) +\int_0^z ({\rm cos}(t) -1)dt/t\,,\qquad {\rm
Si}(z)\equiv
\int_0^z {\rm sin}(t)dt/t\,,$$
$\gamma\simeq 0.577$  is  the Euler-Mascheroni 
constant, and the distance scale $r_c$ is defined as follows:
\beq
r_c \equiv \frac{g^2}{2e^2}\, .
\label{r0}
\eeq
Let us examine more carefully the short- and long-distance
behavior of this expression.

At short distances when $r<<r_c$ we find:
\beq
V(r)~\propto ~{1 \over r}~\left \{
{\pi\over 2} +\left [-1+\gamma+{\rm ln}\left ( {r\over r_c} \right ) 
\right ]\left ( {r\over r_c} \right )~+~{\cal O}(r^2)  
\right \}~.
\label{short}
\eeq
As was expected, at short distances the potential 
has the correct four-dimensional  $1/r$ scaling. 
At intermediate distances it is  modified
by a logarithmic {\it repulsion} term. 
 
Now let us turn to the large-distance behavior. Using (\ref {V})
we obtain for $r\gg r_c$ 
\beq
V(r)~\propto~{1 \over r}~\left \{
{r_c\over r}~+~{\cal O} \left ( {1\over r^2} \right ) 
\right \}~.
\label{long}
\eeq
Thus, the  large-distance potential
scales as $1/r^2$, in full accordance with the five-dimensional  theory
laws.

The physical interpretation of this crossover phenomenon
is as follows. A  gauge field emitted by the source 
localized on the brane propagates along the brane but gradually dissipates 
in  the bulk. The lower the frequency of the 
signal  the faster it leaks in  the extra space 
(this is the phenomenon  of ``infrared transparency''). This is similar 
to what happens with the metastable gravitons in the models of Refs.
\cite {GRS1,Csaki1,DGP1}.
Since in the model at hand the bulk gauge coupling is a free parameter, 
its value  should be fixed by experimental 
bounds on the  photon propagation in four dimensions (see below). 

Returning to the momentum representation we observe
that in Minkowski the propagator (\ref{prop}) develops a discontinuity
across the cut. Denoting the Minkowski (four-)momentum squared by
$p_M^2$ one readily gets both the imaginary and real parts of the
photon propagator on the cut (i.e. at $p_M^2>0$),
\beq
{\rm Re}\, D(p)= \frac{1}{p_M^2 +p_*^2}\,,\qquad
{\rm Im}\, D(p) = \frac{p_*}{p_M}\,\, \frac{1}{p_M^2 +p_*^2}\,,
\label{D}
\eeq
where we suppressed the tensorial structure of $D_{\mu\nu}$. 
 In Sect. 4 we will discuss
  effects due to the deviation of (\ref {D})
from the Feynman propagator at small momenta. Note that
the imaginary part is suppressed with regards to the real part
by one power of ${p_*}/{p_M}$.

\section{Spontaneous Quasilocalization on Finite-Thick\-ness
Domain Wall}

In this section we will consider a
field-theoretic model resulting in a
finite-thickness
domain wall and analyze how   quasilocalization of the gauge field
works in this case. The main emphasis is put on effects
which are invisible in the zero-width limit.

Our starting point is the five-dimensional
 model with scalars, spin-1/2 fermions
and the vector fields in the bulk. The Lagrangian has  the form
\beq
{\cal L} &=&
-{1\over 4g^2}~{\cal F}_{AB}^2+ i {\bar \Psi}\gamma^C(\partial_C
+ i\a_C)\Psi~\nonumber \\ [0.2cm]
&+& {1\over 2}(\partial_\mu \phi)^2~-~{\lambda \over 2} 
(\phi^2-v^2)^2~+~\phi~{\bar\Psi}\Psi\,,
\label{bulk3}
\eeq
where $\phi$ is a real scalar field, and $\Psi$
is a five-dimensional spinor. From the point of view of a 
four-dimensional observer $\Psi$ is a set of two Weyl
spinors, one dotted and one undotted, $\Psi =\{\xi, \bar\eta\}$.
Moreover, $\left(\gamma^5\right)_{5\rm D} = -i\left(\gamma^5\right)_{4\rm
D}$.

The equation of motion for the scalar field $\phi$ has  
a well-known kink solution,
\beq
\phi_0(y) = v\, {\rm tanh} \left (\sqrt{\lambda} vy\right )\, .
\label{kink}
\eeq
The corresponding domain wall 
provides  a toy model for our four-dimensional
 world. 

To begin with, let us put the vector field $\a_C =0$.
There are fermion zero modes
localized on the domain wall, i.e. normalizable solutions of the
equation $ i  \left(\gamma^5\right)_{5\rm D} \partial_5 \Psi
+\phi_0 \Psi =0$. For the wall configuration (\ref{kink})
a normalizable solution exists for $\xi$, while for antikink
a normalizable solution exists for $\eta$, i.e. 
  the zero modes are chiral \cite {JackiwRebbi}. 

From the standpoint of a  four-dimensional observer  the only fermionic
mode which is light is
\beq
\Psi(x,y)=\{ \xi (x) \,   f(y),\,\, 0\}\,,\qquad
f(y)\equiv {\cal N}
{\rm exp} \left( -\int_0^y
\phi_0(z) dz \right )\, ,
\label{zm}
\eeq
where $\xi_\alpha (x)$ is the  four-dimensional fermion field
($\alpha =1,2$); it is left-handed. Moreover, $ {\cal N}$
is a normalization factor introduced in such a way that
$\int dy f^2 (y) =1$.

All other modes
acquire masses of order $\sqrt{\lambda} v$.
In the low-energy approximation these heavy modes
should be integrated out. The result of this procedure is well-known,
it gives rise to  higher-dimensional operators which are 
suppressed by powers of $v$. The only remnant which is important
is a Chern-Simons term which could have been anticipated 
{\em a priori}.  Indeed, while   five-dimensional electrodynamics
of the field $\Psi$ is well defined, this is not the case
for four-dimensional electrodynamics of a chiral field $\xi$;
it is anomalous. The Chern-Simons term eliminates this anomaly.

The  five-dimensional Chern-Simons term  generated in the bulk \cite
{CallanHarvey} has the form
\beq
k ~\epsilon^{ABCDE} {\cal A}_A {\cal F}_{BC} {\cal F}_{DE}\,,
\label{CS}
\eeq
where $k$ is some dimensionless constant. 
This term provides anomaly inflow from the bulk in  the 
 theory living on the wall so  that the axial anomaly due
to the coupling of the localized chiral zero-modes
to the vector field is canceled by the 
surface term emerging from the variation of the 
Chern-Simons term in (\ref  {CS}) \cite {CallanHarvey}\footnote{
In general, one could  
introduce $N_f$ flavors with different charges  
so that  the anomaly in the worldvolume theory is canceled.}.

Having  said this, let us return  to the 
localized massless modes on the wall, and switch on the
gauge field. From the standpoint of a four-dimensional observer
${\cal A}_\mu$  (${\cal A}_y$) is a vector (scalar) field.
The latter is not coupled to the zero mode.
This follows from the fact that the only 4D scalar bilinear one can write
is $\xi_\alpha\xi^\alpha$ and its Hermitian conjugate.
This combination is charged, however, with respect to the conserved U(1)
present in  (\ref{bulk3}) and, thus, cannot appear. Thus, the chirality of
the zero modes prevent nonderivative couplings
of ${\cal A}_y$ to localized matter.

The 
coupling of the zero mode to the  bulk
vector field is as follows:
\beq
\bar\xi_{\dot\alpha} (x) \xi_\alpha (x) (\sigma^\mu )_{\dot\alpha\alpha} \int
dy\, 
\a_\mu (x,y)~f^2(y) \equiv \bar\xi_{\dot\alpha} (x) \xi_\alpha (x)(\sigma^\mu
)_{\dot\alpha\alpha}  \, a_\mu(x) \,,
\label{a}
\eeq
where  we   defined a new four-dimensional
 vector  field $a_\mu (x)$ as $$a_\mu(x) =\int dy\, 
\a_\mu (x,y)~f^2(y).$$
The  four-dimensional chiral fermions are coupled to $a_\mu(x)$
in the standard manner, in the way the charged fermions should be coupled
to the U(1) gauge field.  This
coupling necessarily  induces a  gauge kinetic term for the vector-potential
$a_\mu(x)$, via the fermion loop with two external $a_\mu(x)$ legs, 
much in the same way as in Eq. (\ref{4DF}),
with the substitution of $A_\mu$ of Eq. (\ref{pull}) by  $a_\mu $.
The coefficient in front of this term has the right sign.

The total low-energy action 
for  the fermions and gauge  fields  
is the sum  of the bulk kinetic term (\ref {bulk}), 
the Chern-Simons term (\ref {CS}),
the induced term
\beq
-{1\over {4 e^2}}\, \left(\partial_\mu a_\nu - \partial_\nu a_\mu\right)^2\,,
\label{4DFprim}
\eeq
 and the  interaction term (\ref {a}).
Besides, there is a massless Goldstone boson  and the  massless $\a_y$
field in the world-volume.  The former interacts  with fermions
with derivatives  and
does not contribute to the Coulomb potential between the localized sources.
The latter field  has no relevant interactions with localized fields in the 
given model\footnote{There are interaction vertices of the 
$\a_y$ scalar with one light and one heavy 
state. Thus, $\a_y$ can be produced by a 
light state in high energy processes with the energy of order  
$\sim v$.}.

Let us now study the Coulomb force between 
the localized sources on the brane. This 
problem is quite similar  to the 
one  studied in the previous section.
There are two modes of propagations  of the vector field  between
any two points on the brane. One is a 
bulk kinetic term for the gauge fields and the other one is the induced
kinetic term. As before, choosing the Lorentz gauge in the bulk
we write   equations of motion analogous to (\ref{eqs}),
\beq
&&\partial_C\partial^C \a_{\mu}(x,y) + \frac{g^2}{e^2}  f^2(y) 
\left [ \partial_\nu\partial^\nu
\a_\mu~+~\partial_\mu (\partial_y \a_y)  \right ]~=~g^2 J_\mu(x) \d ~,
\nonumber \\[0.2cm]
&&\left ( \partial_\mu\partial^\mu~+~\partial_y^2 \right ) \a_y~=~0.
\label{eqs3}
\eeq
Making a (partial)
Fourier transformation to the momentum space, as defined in the previous
section, we arrive at    
\beq
(p^2-\partial_y^2)\, {\tilde \a}_\mu(p,y) +\frac{g^2}{e^2} f^2(y) 
 \left [ p^2 ~{\tilde \a}_\mu (p,y) 
+ip_\mu \partial_y {\tilde \a}_y
\right ] = g^2\, {\tilde J}_\mu(p)  \d\,.
\label{eqA}
\eeq
The function $f(y)$ is   peaked and 
quickly drops down to zero  away from the origin.
To solve this equation we approximate it by a step-function {\em  ansatz},
\beq
f(y)= \alpha ~~~{\rm for}~~~ |y|<y_0,~~~f(y)= 0~~~{\rm for}~~~ |y|>y_0,
\label{f}
\eeq
where $y_0$ is some positive constant. The delta-function approximation 
studied in the previous section is obtained in the limit 
\beq
y_0\to 0,~~\alpha \to \infty,~~~{\rm with}~~
y_0\alpha~~ {\rm const.}
\label{limit}
\eeq
Equation  (\ref {eqA}) can be solved piecewise
in  three different intervals. The solutions are
\beq
&&{\tilde \a}_\mu(p,y)=  g^2~{\tilde J}_\mu(p)\left \{   
{1 \over 2\sqrt{\beta}p  }e^{-p\sqrt{\beta}|y|}~+~
b~e^{-p\sqrt{\beta}y}~+~d~e^{p\sqrt{\beta}y} \right \}, 
\nonumber \\[0.2cm]
&&{\rm  for}~~~
|y|<y_0\,,\qquad {\rm where}~~~  \beta \equiv 1+\alpha^2~\frac{g^2}{e^2}\,;
\label{inside}
\eeq
\beq
{\tilde \a}_\mu(p,y)=  g^2\, {\tilde J}_\mu(p) \,c \,
{\rm exp}(-py),~~~{\rm for}
~~~y>y_0 \nonumber \\[0.3cm]
{\tilde \a}_\mu(p,y)=  g^2\,{\tilde J}_\mu(p) 
\,c \,{\rm exp}(py),~~~{\rm for}
~~~y<-y_0 ,
\label{outside}
\eeq
where $b,d$ and $c$ are momentum-dependent constants which can be
determined by matching  the solutions and  their derivatives at $y=\pm y_0$.

In addition, we get that $\partial_y{\tilde \a}_y=0$. 
Our primary  interest is  the solution confined to the wall.
This is determined by Eq. (\ref {inside}) near the point $y=0$,
\beq
{\tilde \a}_\mu(p,0)=g^2~{\tilde J}_\mu(p) 
\left \{   
{1 \over 2 \sqrt{\beta} p  } +{1\over 2\sqrt{\beta}p}
 { (\sqrt{\beta}-1) {\rm exp}(-\sqrt{\beta}y_0p)\over
{\rm ch}(\sqrt{\beta}y_0p ) ~+~\sqrt{\beta}~{\rm sh}(\sqrt{\beta}y_0p) }
\right \}.
\label{in}
\eeq
Taking the limit (\ref {limit})
we find 
\beq
{\tilde \a}_\mu(p,y=0)~=~ {{\tilde J}_\mu(p)\over 
p^2+pp_*}\,,
\eeq
i.e.  precisely the expression which we analyzed in the previous
section. Thus, the vector fields are quasilocalized due to the 
induced kinetic term on the domain wall.

\section{Some Phenomenological and Astrophysical Implications}

Let us first discuss the mass scale relevant to   five- and 
four-dimensional
physics. The mass parameter relevant to five-dimensional
physics is $g^{-2}$. Since 
the five-dimensional theory under consideration is an
effective one, a  five-dimensional cut-off $\Lambda_5$
must be introduced.  The loop expansion in the
 five-dimensional theory runs in powers of
${g^{2}\Lambda_5 }/{4\pi }\,. $
For the theory to make sense we must require that
$ \Lambda_5 \sim g^{-2}\,.$
On the other hand, the scale of $g^{-2}$ is set from   requiring
  deviations from the Coulomb law in our wall-confined world
to be acceptable, i.e. parameter $p_*$ be small enough.
Barring the possibility of exceedingly large $N_f$ one can say that
$p_*$ and $g^{-2}$ are of the same order of magnitude.

Note that the  four-dimensional world the expansion runs in
$p_*/M = 1/(g^2 M)$ where the parameter $M$ is related to the masses
of the localized modes. In this way this is the strong coupling expansion with
respect to the five-dimensional
theory. However, the above constraints are 
evaded if one adopts a scenario where
the  charged matter fields are confined to  a brane and do not
propagate in the bulk.  Then, in the case of the Abelian gauge field,
there are no loops in the bulk, all loops are confined to the
wall world-volume\footnote{Another option, which we will not pursue here,
is to make $g^{2} $ defined by a scalar field (``dilaton'')
which has a $y$ dependent profile so that $g^{2}$ is
large on the brane while it is suppressed away from the brane.}.

Let us discuss some bounds on the crossover scale $r_c$.
At  distances $r > r_c$ the electrodynamics 
we consider becomes
five-dimensional. At   first sight, the bound on $r_c$ is expected
to be very severe, at least comparable to the present Hubble size. 
This is due to the
fact that we are constantly detecting electromagnetic 
waves propagating
over the cosmic distances, and these waves seem to behave
in a perfectly four-dimensional way. 

Surprisingly enough, the actual bound on $r_c$ is rather mild, 
as we   explain below.
This is due to the phenomenon of an ``infrared transparency'' 
according to which the large wavelength radiation penetrates easier
in   extra dimensions.
To illustrate the point let us consider an electromagnetic wave produced
by a monochromatic source $l$ located on the brane
\beq
J_{\mu}(x,y)~\sim~l_{\mu}~ \delta(y)~\delta^{(3)}(x)~{\rm exp}(i\omega t)
\,.
\label{l}
\eeq
The corresponding wave equation is given in (\ref {eqs3})
where the right-hand side is substituted by (\ref {l})
and $\a_y$ is put to zero. 
Qualitatively, the wave behavior   can be understood as 
follows. 
If it were  not for the brane world-volume contribution
(the second term on the left-hand side of Eq.~(\ref {eqs3}))
the wave would behave as  five-dimensional,
\beq
 \sim ~\sqrt {\omega}~{e^{i\omega(t-R)} \over R^{3\over 2} }\,.
\label{5DW}
\eeq
where $R$ stands for the five-dimensional radial coordinate. 
On the other hand, a four-dimensional wave which propagates in the 
world-volume would be described by 
\beq
 \sim {e^{i\omega(t-R)} \over R}~\d~.
\label{4DW}
\eeq
Using these expressions and Eq. (\ref {eqs3})
one can estimate a distance in the brane world-volume 
at which the crossover
between the  four-dimensional and five-dimensional behavior should take
place,
\beq
r ~\sim~ r_{\omega}~\sim~\omega r_c^2~.
\label{ro}
\eeq
Thus $r_{\omega}$ sets the distance below which the wave propagates as
  four-dimensional. Surprisingly enough, this resembles 
the behavior of the gravitational waves in \cite{GRS1}. 
For short  wave length  physics becomes exceedingly 
 more four-dimensional.
The waves with the  frequency
$\omega \gg  1/r_c$ will propagate as four-dimensional waves 
over the distances much
larger than $r_c$.  This suggests
that even if the Coulomb law
gets modified at relatively short distances, propagation
of  visible light at larger scales will still look perfectly
four-dimensional.
For instance, let us assume that the Coulomb law breaks down beyond the
solar system size, that is to say $r_c \sim 10^{15}$ cm. 
The largest wave-length radiation
propagating over cosmic distances which  has been detected so far
are the radio waves in a meter wave-length range. Such radiation would
propagate according to the laws of four-dimensional physics 
over the  distance scale 
\beq
 r_{\rm radio} \sim 10^{28}~ {\rm cm}~.
\eeq
This is comparable to the size of the Universe. Thus, the Coulomb law
might break down at a scale of
 the solar system size  and we would 
not even notice it!

By the same reason, astrophysical processes, such as  star cooling,
cannot provide any significant bound on $r_c$. 
The photons emitted by   stars will remain
four-dimensional over the scales larger that the present Hubble distance.
In this respect, the radio astronomy may provide a  useful bound on
$r_c$ through precision measurements of  the 
frequency-on-intensity dependence of the large wave-length radiation. 

Another prediction of the present scenario, is a possibility
of the charge non-conservation
on the brane.  Non-conservation of global charges is a generic feature of
brane world scenarios \cite {DGGlobal}. However, local charges, 
the ones that  are associated with the unbroken local  symmetries
are expected to be conserved in the world-volume due to the flux
conservation  arguments. This takes place   
in the scenario
of Ref. \cite{GMLoc} where  photon is strictly localized on the brane
due to the condensation of magnetic charges in the bulk. 
As a result, the electric flux is repelled from the exterior and spreads
on the brane, producing a four-dimensional Coulomb potential between
the two brane-localized sources.
In such a scenario, if one tries to remove a  $Q$-amount of the electric 
charge from the brane,
the flux tube
will be stretched between the brane and the particle. The throat of the
flux tube on the   brane will be seen by the brane observer as a charged
particle of the same charge $Q$  and no 
apparent charge violation will be present.
The flux tube may break into  charge-anticharge pairs, and neutral
states can escape into    extra dimensions. However, the end of the flux
tube that will remain attached to the brane will still carry the same charge 
$Q$.

In our present scenario photons may leak in  extra dimensions, and
thus, electric charge
can leak there too. Both of these effects give new striking features,
very different from the standard brane-world scenarios in which gauge
fields are localized on the brane. 

One may think of various astrophysical implications of the photon
disappearance in  extra dimensions. For instance, there is a certain
observational evidence \cite {supernovae} that distant
supernovae appear to be
dimmer than it is  expected in the standard matter-dominated
cosmological scenario.  
The results of these studies show that the distant supernovae are
fainter. In the conventional four-dimensional
 theory this would mean that
they are more distant than expected for the decelerating Universe, and,
 thus,
indicate that the Universe is accelerating. In our case, however,
this may mean that photons spread to extra dimensions. 
   
A natural question in our framework is whether 
or not the dimmer
supernovae can be  simply the consequence of the fact that photons emitted
by these stars dissipate into the fifth dimension  on their way to the
Earth? Unfortunately, this idea is not very easy to realize. 
For instance, the
immediate question would be: What prevents the cosmic microwave background
radiation (CMBR) from dissipating into the extra dimension? 
(The CMBR  wave-length is much
longer than that for the radiation detected  
from distant supernovae\footnote{We
thank V.~Rubakov for commenting on this issue.}.)
In this respect it may be interesting to find  a scenario 
in which the resonance photon  
has a decay width which vanishes below 
some frequency $\omega_0$ and is nonzero for 
higher frequencies. Assuming that  
$\omega_0$ is less than  the visible light
frequency ($10^{6}$ m$^{-1}$), 
but is above the frequency of CMBR 
(mm$^{-1}$), the CMBR photons would not  be able to  escape in
the bulk. On the other hand, the  photons
emitted by the distant supernovae are more energetic and would dissipate
in  extra dimensions.  Of course, even in this case it
is not at all clear that the frequency dependence on the disappearance
rate can correctly fit into  the observed supernovae spectrum, so that 
a more careful study is required. 

Another comment concerns 
the galactic magnetic fields.
These are used to put constraints on photon mass.
In our case, however, we deal with  a mild power-law  
modification of the Coulomb
law and the use of galactic magnetic 
fields to constraint $r_c$  would only be possible if 
detailed properties  and  the origin  
of these fields were known. 
Unfortunately, this information  is not 
available at present (see, e.g., \cite {Rubin}).

\section{A Byproduct Application?}

One may try to exploit the very same idea of (quasi)localization of the
gauge fields within the framework of Kaplan's suggestion
\cite{Kaplan} to generate chiral fermions on the lattice
by considering five-dimensional lattice with a domain wall
to which fermion zero modes  are confined. 
As long as these wall-confined fermions carry 
color quantum numbers, they will generate, through the loops
a kinetic term to the gauge fields which will be peaked on the
four-dimensional wall. If the lattice parameters are appropriately chosen,
the above kinetic term may force localization of the gauge fields
on the same wall.

\vspace{0.2in}

{\bf Acknowledgments}

\vspace{0.1in}

The authors are grateful to B. Bajc, C. Deffayet, 
A. Glassgold, P. Huggins, V. Rubakov and 
A. Sirlin for useful  discussions. G.D. and G.G. thank the {\it Aspen
Center for  Physics} for hospitality where a part of this project was 
performed. The work of G.D. was supported by David and Lucille 
Packard Foundation 
Fellowship for Science and Engineering, 
and by Alfred P. Sloan Foundation 
Fellowship and by NSF grant PHY-0070787. Work of G.G. and M.S. was 
supported by DOE grant DE-FG02-94ER408.


\begin{thebibliography}{99}

\bibitem{ADD} N. Arkani-Hamed, S. Dimopoulos, G. Dvali,
Phys. Lett. {\bf B429}, 263 (1998); 
Phys. Rev. {\bf D59}, 0860 (1999); 
I. Antoniadis,  N. Arkani-Hamed, S. Dimopoulos, G. Dvali,
Phys. Lett. {\bf B436}, 257 (1998)~.

\bibitem{RandallSundrum}  L. Randall, R.  Sundrum,
Phys. Rev. Lett. {\bf 83}, 3370  (1999)~; 
{\it ibid.} {\bf83}, 4690 (1999)~.

\bibitem{GiaMisha} 
G. Dvali, M. Shifman, Nucl. Phys. {\bf B504}, 127  (1997)
[hep-th/9611213]. 

\bibitem{FAN}
G.~Dvali and M.~Shifman,
Phys.\ Lett.\  {\bf B475}, 295 (2000)
[hep-ph/0001072].

\bibitem{RubakovShaposhnikov}  
V.~A.~Rubakov and M.~E.~Shaposhnikov,
Phys.\ Lett.\  {\bf B125}, 136 (1983).

\bibitem{JackiwRebbi} 
R.~Jackiw and C.~Rebbi,
Phys.\ Rev.\  {\bf D13}, 3398 (1976).

\bibitem{BorutGiga}
B.~Bajc and G.~Gabadadze,
Phys.\ Lett.\  {\bf B474}, 282 (2000)
[hep-th/9912232].

\bibitem{RandjbarShaposhnikov} 
S.~Randjbar-Daemi and M.~Shaposhnikov,
hep-th/0008079.

\bibitem{Polchinski} 
J. Polchinski, Phys. Rev. Lett. {\bf 75}, 4724 (1995).

\bibitem{GMLoc} 
G. Dvali, M. Shifman, Phys.\ Lett.\  {\bf B396}, 64 (1997);
Erratum {\bf B407}, 452 (1997) [hep-th/9612128].

\bibitem{Oda} 
I.~Oda,
hep-th/0006203.


\bibitem{Tinyakov} 
S.~L.~Dubovsky, V.~A.~Rubakov and P.~G.~Tinyakov,
JHEP {\bf 0008}, 041 (2000).

\bibitem{DGPind} 
G.~Dvali, G.~Gabadadze and M.~Porrati,
Phys.\ Lett.\  {\bf B485}, 208 (2000).

\bibitem{DG} G. Dvali, G. Gabadadze, 
hep-th/0008054.


\bibitem{GRS1} R.~Gregory, V.~A.~Rubakov, S.~M.~Sibiryakov,
Phys.\ Rev.\ Lett.\  {\bf 84}, 5928 (2000).

\bibitem{Csaki1}
C.~Csaki, J.~Erlich and T.~J.~Hollowood,
Phys.\ Rev.\ Lett.\  {\bf 84}, 5932 (2000).

\bibitem{DGP1} 
G.~Dvali, G.~Gabadadze and M.~Porrati,
Phys.\ Lett.\  {\bf B484}, 112 (2000); 
Phys.\ Lett.\  {\bf B484}, 129 (2000).



\bibitem{CallanHarvey} 
C.~G.~Callan and J.~A.~Harvey,
Nucl.\ Phys.\  {\bf B250}, 427 (1985).

\bibitem{DGGlobal}
G.~Dvali and G.~Gabadadze,
Phys.\ Lett.\  {\bf B460}, 47 (1999)
[hep-ph/9904221].

\bibitem{supernovae}  M. Hamuy, et al., Astroph. Journ. {\bf 109} 
1, (1995); ibid {\bf 112}, 2391 (1996); \\
A.G. Riess, et al., Astrop. Journ.,  {\bf 117} 707 (1999).       

\bibitem{Rubin} D. Grasso, H.R. Rubinstein, astro-hep/0009061~.

\bibitem{Kaplan}
D.B. Kaplan,  Phys. \ Lett. \ {\bf B288}, 342 (1992) 
[hep-lat/9206013]. 

\end{thebibliography}
\end{document}